\documentclass[11pt,leqno]{article}

\usepackage{mathptmx}

\usepackage{amssymb,amsmath,amsthm}
\usepackage{comment}
\usepackage[latin1]{inputenc}    
\usepackage{pifont}    

\makeatletter%

\usepackage{nicefrac}
\usepackage{mathptmx}
\newcommand{\doublespacing}{\let\CS=
\@currsize\renewcommand{\baselinestretch}{1.75}\tiny\CS}
\newcommand{\extradoublespacing}{\let\CS=
\@currsize\renewcommand{\baselinestretch}{1.9}\tiny\CS}
\newcommand{\draftspacing}{\let\CS=
\@currsize\renewcommand{\baselinestretch}{2.0}\tiny\CS}
\newcommand{\hugedraftspacing}{\let\CS=
\@currsize\renewcommand{\baselinestretch}{2.4}\tiny\CS}
\makeatother%

\newcommand{\OMIT}[1]{} %
\newcommand{\gfootnote}[1]{} %

\newcommand\qedblob{\ding{113}}
\def\literalqed{{\ \nolinebreak\hfill\mbox{\qedblob\quad}}}

\newenvironment{proofs}{\noindent{\bf Proof.}\hspace*{1em}}{\literalqed\bigskip}

\newcommand{\scoreof}[1]{\mathit{score}(#1)}
\newcommand{\scoresub}[2]{\mathit{score}_{#1}(#2)}

\newcommand{\scoresublevel}[3]{\mathit{score}_{#1}^{#2}(#3)}

\newcommand{\wtwo}{{\ensuremath{{\rm{W}}[2]}}}
\newcommand{\wtwobf}{{\ensuremath{{\rm{\bf{W}}}[\bf{2}]}}}

\newcommand{\p}{\mbox{\rm P}}

\newcommand{\np}{\mbox{\rm NP}}

\newcommand{\condition}{\,|\:}

\newenvironment{desctight}
  {\begin{list}{}{\setlength\labelwidth{0pt}%
        \setlength{\itemsep}{0.5pt}%
        \setlength{\parsep}{0pt}%
        \setlength\itemindent{-\leftmargin}%
        }}
    {\end{list}}

  \newtheorem{theorem}{Theorem}[section]

  \newtheorem{definition}[theorem]{Definition}

\usepackage{ulem}
\begin{document}

\title{Parameterized Control Complexity in Fallback
Voting\thanks{This work was supported in part by the DFG under grants 
RO~\mbox{1202/12-1} (within the European Science
Foundation's EUROCORES program LogICCC: 
``Computational Foundations of Social Choice'') and RO~\mbox{1202/11-1}.  
  Work done in part while the first author was visiting 
  the University of Newcastle.}
}
\author{
G\'{a}bor Erd\'{e}lyi\thanks{URL: 
ccc.cs.uni-duesseldorf.de/\symbol{126}erdelyi.
Heinrich-Heine-Universit\"{a}t D\"{u}sseldorf,
Institut f\"{u}r Informatik, 
40225 D\"{u}sseldorf,
Germany.
}
\quad and \quad
Michael Fellows\thanks{URL: http://mrfellows.net/.
University of Newcastle, %
Callahan, 
NSW Australia 2308.
}
}

\date{April 21, 2010}

\maketitle

\begin{abstract}
 We study the parameterized control complexity of fallback voting, a voting system that
combines preference-based with approval voting. 
Electoral control is one of many different ways for an external agent to tamper with the 
outcome of an election. We show that adding and deleting candidates in fallback voting  are 
$\wtwo$-hard for both the constructive and destructive case, parameterized by the amount of 
action taken by the external agent. Furthermore, we show that adding and deleting voters
in fallback voting are $\wtwo$-hard for the constructive case, parameterized by the amount of 
action taken by the external agent, and are in FPT for the destructive case.
\end{abstract}
\maketitle                   %

\section{Introduction}
 \label{sec:introduction}
The study of algorithmic issues related to voting systems has moved front-and-center
in contemporary computer science.  Google is basically an algorithmic engine (plus
targeted advertising) that collates ranking information (votes) mined from series-of-clicks
and other data, tuned by further heuristics. 
Essentially, every Google query result is the outcome of an algorithmic 
{\it election} concerning the
relevance of websites to the query.

Similar issues arise throughout science, in the current era of vastly expanded pools of raw
information.  Many of these issues can be thought of as specialized Google-style queries: for example, ``which genes in the database(s) seem to be most relevant to this new information
...?''

In all areas of science, government and industry, the collation 
of information, and prioritization of allied strategic options, has moved front-and-center
both tactically and strategically.

This general situation has profoundly stimulated, and drawn upon, a mathematical subject that used to be a bit of a backwater, concerned with ``ideal'' voting systems and so forth.  But now
(and forever more) rapid and frequent, algorithmically-powered, elections about relevant information are becoming a
cornerstone of civilization.  

As in (almost) all things algorithmic, rich questions 
inevitably arise about the tractability of the desired information-election processes, and their susceptibility 
to manipulation (or primary data error).  This paper is about this context of research.

\section{Preliminaries}
\label{sec:preliminaries}

Given the many information-resolution contexts in which voting systems are relevant, 
presenting different characteristics and challenges, it is an important 
resource that various voting systems have been proposed.  
These are now being vigorously
investigated in regards algorithms and complexity issues of their strengths and weaknesses
in various applied contexts.
There are many papers regarding the complexity-theoretic aspects of the many different ways of changing 
the outcome of an election, like \emph{manipulation} 
\cite{bar-tov-tri:j:manipulating,bar-orl:j:polsci:strategic-voting,con-san-lan:j:when-hard-to-manipulate,hem-hem:j:dichotomy-scoring,fal-hem-hem-rot:c:single-peaked-preferences}, where a group of voters cast their votes
strategically, \emph{bribery}
\cite{fal-hem-hem:j:bribery,fal-hem-hem-rot:j:llull-copeland-full-techreport}, where an
external agents bribes a group of voters in order to change their votes, and \emph{control}
\cite{bar-tov-tri:j:control,hem-hem-rot:j:destructive-control,fal-hem-hem-rot:j:llull-copeland-full-techreport,hem-hem-rot:j:hybrid,erd-now-rot:j:sp-av,fal-hem-hem-rot:c:single-peaked-preferences,erd-pir-rot:t-With-Complete-CATS10-Ptr:fallback-voting},
where an external agent---which is referred to as ``The Chair''---changes the structure of the election (for example, by 
adding/deleting/partitioning either candidates or voters).

In this paper, we are concerned with {\it control issues} for the relatively recently introduced voting system of {\it fallback voting} (FV, for short)~\cite{bra-san:j:preference-approval-voting}.  
Fallback voting is the natural voting system that currently has the most resistances (i.e., makes the chair's task hard) for control attacks
\cite{erd-pir-rot:t-With-Complete-CATS10-Ptr:fallback-voting}. We investigate the issues in the framework 
parameterized complexity.  Many voting systems present NP-hard algorithmic challenges.  Parameterized complexity is a particularly appropriate framework in many contexts of voting
systems because it is concerned with exact results that exploit the structure of input distributions.  It is not appropriate in political contexts, for example, to 
algorithmically determine a winner ``approximately''. 

In this section, we explicate voting systems in general, fallback voting in particular,
parameterized complexity theory, and
some graph theory that we will use.

\subsection{Elections and Electoral Control}

An election $(C,V)$ consists of a finite set of candidates $C$ 
and a finite collection of voters $V$ who express their preferences 
over the candidates in $C$, and distinct voters can have the same 
preferences.
A voting system is a set of rules determining the winners of an 
election. In our paper we only consider the unique-winner model, where
we want to have exactly one winner at the time. Votes can be represented 
in different ways, depending on the voting system used. 
One widely-used representations of votes is via preference rankings.
In this case each voter has to specify a tie-free linear ordering of all candidates.
Such voting systems are for example Condorcet, Borda count, plurality or
veto; see, e.g., \cite{bra-fis:b:voting-procedures}. 
Approval voting, introduced by Brams and Fishburn 
\cite{bra-fis:j:approval-voting,bra-fis:b:approval-voting} is not a 
preference based voting system. In \emph{approval voting} each voter has to 
vote ``yes'' or ``no'' for each candidate and the candidates with the 
most ``yes'' votes are the winners of the election. 
Clearly, approval voting completely ignores preference rankings.

Brams and Sanver \cite{bra-san:j:preference-approval-voting} introduced
two voting systems that combine preference-based with approval voting.
One of these systems is fallback voting.

\begin{definition}[\cite{bra-san:j:preference-approval-voting}]
\label{def:fv}
Let $(C,V)$ be an election.
Every voter $v \in V$ has to divide the set of candidates $C$ into two 
subsets $S_v \subseteq C$ indicating that $v$ approves of all candidates 
in $S_v$ and disapproves of all candidates in $C-S_v$.  $S_v$ is called 
\emph{$v$'s approval strategy}.  In addition, each voter $v\in V$ provides 
also a tie-free linear ordering of all candidates in $S_v$.

Representation of votes:
Let $S_v = \{c_1, c_2,\ldots , c_k\}$ for a voter $v$ who ranks the candidates 
in $S_v$ as follows. $c_1 > c_2 > \cdots > c_k$, where $c_1$ is $v$'s most
preferred candidate and $c_k$ is $v$'s least preferred candidate. We denote the 
vote $v$ by
\[
\begin{array}{c@{\ \ }c@{\ \ }c@{\ \ }c@{\ \ }c@{\ \ }c}
 c_1 & c_2 & \cdots & c_k & | & C - S_v,
\end{array}
\]
where the approved candidates to the left of the approval line are
ranked from left to the right and the disapproved candidates to the
right of the approval line are not ranked and written as a set $C-S_v$.
In our constructions, we sometimes also insert a subset $B\subseteq C$
into such approval strategies, where we assume some arbitrary, fixed order
of the candidates in $B$ (e.g., ``$ \begin{array}{c@{\ \ }c@{\ \ }c@{\ \ }c}
 c_1 & B & | & (C - B - \{c_1\})
\end{array}$'' means that $c_1$ and all $b\in B$ are approved of, while the
rest of the candidates are disapproved of).

Let $\scoresub{(C,V)}{c} = \|\{v \in V \condition c \in S_v\}\|$
denote the number of voters who approve of candidate $c$, and let
$\scoresublevel{(C,V)}{i}{c}$ be the \emph{level~$i$ score of $c$ in
  $(C,V)$}, which is the number of $c$'s approvals when ranked between
the (inclusively) first and $i^{th}$ position.

Winner determination:
\begin{enumerate} 
 \item On the first level, only the highest ranked approved candidates
   (if they exist) are considered in each voters' approval strategy. 
   If there is a candidate $c\in C$ with $\scoresublevel{(C,V)}{1}{c} > \nicefrac{\|V\|}{2}$
   (i.e., $c\in C$ has a strict majority of approvals on this level), 
   then $c$ is the \emph{(unique) level~$1$ FV winner of the election},
   and the procedure stops.
 
\item If there is no level~$1$ winner, we ''fall back`` to the second
  level, where the two highest ranked approved candidates (if they exist) 
  are considered in each voters' approval strategy. If there is exactly 
  one candidate $c\in C$ with $\scoresublevel{(C,V)}{2}{c} > \nicefrac{\|V\|}{2}$, 
  then $c$ is the \emph{(unique) level~$2$ FV winner of the election},
  and the procedure stops.  
  If there are at least two such candidates, then every candidate with the
  highest level~$2$ score is a \emph{level~$2$ FV winner of the election},
  and the procedure stops.
 
\item If we haven't found a level~$1$ or level~$2$ FV winner, we in 
  this way continue level by level until there is at least one candidate 
  $c\in C$ on a level $i$ with $\scoresublevel{(C,V)}{i}{c} > \nicefrac{\|V\|}{2}$, 
  If there is only one such candidate, he or she is the
  \emph{(unique) level~$i$ FV winner of the election},
  and the procedure stops.  If there are
  at least two such candidates, then every candidate with the highest
  level~$i$ score is a \emph{level~$i$ FV winner of the election},
  and the procedure stops.
 
 \item If for no $i \leq \|C\|$ there is a level~$i$ FV winner, every candidate 
   with the highest $\scoresub{(C,V)}{c}$ is a \emph{FV winner of~$(C,V)$ by score}.
\end{enumerate}
\end{definition}

We now formally define the computational problems that we 
study in our paper. We consider two different control types. 
In a \emph{constructive} control scenario, introduced by Bartholdi, Tovey,
and Trick \cite{bar-tov-tri:j:control}, the chair seeks to make his
or her favourite candidate win the election.
In a \emph{destructive} control scenario, introduced by Hemaspaandra,
Hemaspaandra, and Rothe \cite{hem-hem-rot:j:destructive-control}, 
the chair's goal is to prevent a despised candidate from winning
the election. We will only state the constructive cases. The questions
in the destructive cases can be asked similarly with the difference
that we want the distinguished candidate \emph{not to be} a unique 
winner.

We first define control via adding a limited number of candidates.

\begin{desctight}

\item[Name] Control by Adding a Limited Number of Candidates.

\item[Instance] An election $(C\cup D,V)$, where $C$ is the set of qualified 
candidates and $D$ is the set of spoiler candidates, a designated candidate 
$c\in C$, and a positive integer $k$.

\item[Parameter] $k$.

\item[Question] Is it possible to choose a subset
$D'\subseteq D$ with $||D'||\leq k$ such that $c$ is the unique 
winner of election $(C\cup D', V)$?
\end{desctight}

In the following control scenario, the chair seeks to reach his or her 
goal by deleting (up to a given number of) candidates.

\begin{desctight}

\item[Name] Control by Deleting Candidates.

\item[Instance] An election $(C,V)$, a designated candidate $c\in C$, and a 
positive integer $k$.

\item[Parameter] $k$.

\item[Question] Is it possible to delete up to $k$
candidates (other than $c$) from $C$ such that $c$ is the unique winner 
of the resulting election?
\end{desctight}

Turning to voter control, we first specify the problem control by 
adding voters.

\begin{desctight}

\item[Name] Control by Adding Voters.

\item[Instance] An election $(C,V\cup W)$, where $V$ is the set of registered 
voters and $W$ is the set of unregistered voters, a designated candidate 
$c\in C$, and a positive integer $k$.

\item[Parameter] $k$.

\item[Question] Is it possible to choose a subset
$W'\subseteq W$ with $||W'||\leq k$ such that $c$ is the unique 
winner of election $(C, V\cup W')$?
\end{desctight}

Finally, the last problem we consider, control by deleting voters.

\begin{desctight}

\item[Name] Control by Deleting Voters.

\item[Instance] An election $(C,V)$, a designated candidate $c\in C$, and a 
positive integer $k$.

\item[Parameter] $k$.

\item[Question] Is it possible to delete up to $k$
voters from $V$ such that $c$ is the unique winner of the resulting 
election?
\end{desctight}

The above defined problems are all natural problems, see the discussions
in \cite{bau-erd-hem-hem-rot:t:computational-apects-of-approval-voting,bar-tov-tri:j:control,hem-hem-rot:j:destructive-control,fal-hem-hem-rot:j:llull-copeland-full-techreport,hem-hem-rot:j:hybrid}.

\subsection{Parameterized Complexity}

The theory of parameterized complexity 
offers toolkits for two tasks: (1) the fine-grained analysis of the sources of 
the computational complexity of NP-hard problems, according to secondary measurements (the {\it parameter}) of problem inputs (apart from the overall
input size $n$), and (2) algorithmic methods for exploiting parameters that contribute favorably to problem complexity.  Formally, a parameterized decision problem is a language
${\mathcal L} \subseteq \Sigma^{*} \times N $.  ${\mathcal L}$ is {\it fixed-parameter tractable} (FPT) if and only if it can be determined, for input $(x,k)$ of size 
$n=|(x,k)|$, whether $(x,k) \in {\mathcal L}$ in time $O(f(k)n^{c})$.  This central idea
of parameterized complexity supports a notion of FPT-reducibility that exposes evidence
of likely parameterized intractability, through parameterized reducibility from problems
that are hard or complete for intractable parameterized classes. 
The main hierarchy of parameterized complexity classes is
$$ FPT \subseteq W[1] \subseteq W[2] \subseteq \cdots \subseteq W[P] \subseteq XP. $$
$W[1]$ is a strong analog of NP, as the $k$-Step Halting Problem for Nondeterministic Turing Machines is complete for $W[1]$. 
The $k$-Clique problem is complete for $W[1]$, and the
parameterized Dominating Set problem is complete for $W[2]$. 
See the Downey-Fellows~\cite{dow-fel:b:parameterized-complexity} monograph for further background.

\subsection{Graphs}
\label{sec:preliminaries:graphs}

Many problems proven to be $\wtwo$-hard are derived from problems concerning 
graphs. We will prove $\wtwo$-hardness via parameterized reduction from the problem 
Dominating Set, which was proved to be $\wtwo$-complete by Downey and 
Fellows~\cite{dow-fel:b:parameterized-complexity}. Before the formal 
definition of the Dominating Set problem, we first have to present some
basic notions from graph theory.

An \emph{undirected graph} $G$ is a pair $G=(V,E)$, where $V=\{v_1,\ldots ,v_n\}$ 
is a finite (nonempty) set of vertices and $E=\{\{v_i, v_j\}\condition 1\leq i<j\leq n\}$
is the set of edges.\footnote{In this paper we will use the symbol $V$ strictly for voters.
From the next section on, we will use the symbol $B$ instead of $V$ for the set of 
vertices in a graph $G$.} 
Any two vertices connected by an edge are called \emph{adjacent}.
The vertices adjacent to a vertex $v$ are called the \emph{neighbours} of $v$, and
the set of all neighbours of $v$ is denoted by $N[v]$ (i.e., $N[v]=\{ u\in V\condition
\{u,v\}\in E \}$). The \emph{closed neighbourhood} of $v$ is defined as 
$N_c[v]=N[v]\cup \{v\}$. The parameterized version of Dominating Set is defined as follows.

\begin{desctight}

\item[Name] Dominating Set.

\item[Instance] A graph $G=(V,E)$, where $V$ is the set of vertices and $E$
is the set of edges.

\item[Parameter] A positive integer $k$.

\item[Question] Does $G$ have a dominating set of size $k$ (i.e., a subset 
$V'\subseteq V$ with $||V'||\leq k$ such that for all $u\in V-V'$ there is
a $v\in V'$ such that $\{u,v\}\in E$)?
\end{desctight}

\section{Results}
\label{sec:results}

\begin{table*}[t!]
\centering
{\small
\begin{tabular}{|l||l|l||l|l|}
\hline
                    & \multicolumn{2}{c||}{Classical Complexity}
		    & \multicolumn{2}{c|}{Parameterized Complexity}
\\ \cline{2-5}
Control by          & Constructive & Destructive
		    & Constructive & Destructive

\\ \hline
Adding a Limited Number of Candidates
                    & $\np$-compl.            & $\np$-compl. 
                    & {\bf $\wtwobf$-hard}            & {\bf $\wtwobf$-hard}
\\ \hline
Deleting Candidates & $\np$-compl.            & $\np$-compl.
                    & {\bf $\wtwobf$-hard}            & {\bf $\wtwobf$-hard} 
\\ \hline
Adding Voters       & $\np$-compl.           & $\p$
                    & {\bf $\wtwobf$-hard}            & {\bf (FPT)}

\\ \hline
Deleting Voters     & $\np$-compl.            & $\p$
                    & {\bf $\wtwobf$-hard}            & {\bf (FPT)}
\\ 
\hline
\end{tabular}
}
\caption{\label{tab:summary-of-results}
Overview of results. The classical results are due to Erd\'{e}lyi et al. \cite{erd-rot:c:fallback-voting}.
Results new to this paper are in boldface. 
} 
\end{table*}

Table~\ref{tab:summary-of-results} shows the results on the control
complexity of fallback voting due to Erd\'{e}lyi et al. \cite{erd-rot:c:fallback-voting}
and the new results on the parameterized control complexity of
fallback voting. The FPT results in Table~\ref{tab:summary-of-results}
are in parenthesis because these two results are trivially inherited from
the classical $\p$ results.

In all 
of our results
we will prove $\wtwo$-hardness by 
parameterized reduction from the $\wtwo$-complete problem Dominating 
Set defined in Section~\ref{sec:preliminaries:graphs}. 
In these six proofs we will always start from a given Dominating Set
instance $(G=(B,E),k)$, where $B=\{b_1, b_2, \ldots ,b_n\}$ is the set of 
vertices, $E$ the set of edges in graph $G$, and $k\leq n$ is a positive integer.
In the following constructions, the set of candidates will always contain the
set $B$ which means that for each vertex $b_i\in B$ we will have a candidate
$b_i$ in our election. We will also refer to candidate set $N_c[b_i]$, which is the set of
candidates corresponding to the vertices in $G$ that are in $N_c[b_i]$.

\subsection{Candidate Control}
\label{sec:results:candidate-control}

\begin{theorem}
 \label{thm:adding-candidates}
 Both constructive and destructive control by adding candidates 
in fallback voting are $\wtwo$-hard.
\end{theorem}

\begin{proofs}
 We first prove $\wtwo$-hardness to constructive control by adding
candidates. 
Let $(G=(B,E),k)$ be a given instance 
of Dominating Set as described above.
Define the election $(C,V)$, where $C=\{c,w\} \cup B \cup X \cup Y \cup Z$
with $X=\{x_1, x_2, \ldots , x_{n-1}\}$, 
$Y=\{ y_1, y_2, \ldots ,y_{n-2}\}$, $Z=\{z_1, z_2,\ldots ,z_{n-1}\}$ is the 
set of candidates, $w$ is the distinguished candidate, and $V$ is the following
collection of $2n+1$ voters:

\begin{enumerate}
\item For each $i$, $1\leq i \leq n$, there is one voter of the form:
\[
\begin{array}{c@{\ \ }c@{\ \ }c@{\ \ }c@{\ \ }c}
N_c[b_i] & X & c & | & (B-N_c[b_i]) \cup Y \cup Z \cup \{w\}.
\end{array}
\]
\item There are $n$ voters of the form:
\[
\begin{array}{c@{\ \ }c@{\ \ }c@{\ \ }c@{\ \ }c}
Y & c & w & | & B\cup X\cup Z.
\end{array}
\]

\item There is one voter of the form:
\[
\begin{array}{c@{\ \ }c@{\ \ }c@{\ \ }c}
Z & w & | & B\cup X\cup Y \cup \{c\}.
\end{array}
\]
\end{enumerate}

Note that candidate $w$ is not the unique winner of the election $(C-B,V)$, since
only candidates $c$ and $w$ have a strict majority of approvals, $w$ gets only 
approvals on level $n$, and  $\scoresublevel{(C-B,V)}{n}{w}=n+1<2n=\scoresublevel{(C-B,V)}{n}{c}$
thus, $c$ is the unique FV winner of the election $(C-B,V)$. Now, let $C-B$ be the 
set of qualified candidates and let $B$ be the set of spoiler candidates.

We claim that $G$ has a dominating set of size $k$ if and only if $w$ can be made a
unique FV winner by adding at most $k$ candidates.

From left to right: Suppose $G$ has a dominating set of size $k$. Add the corresponding
candidates to the election. Now candidate $c$ gets pushed at least one position to the 
right in each of the $n$ votes in the first voter group. Thus, candidate $w$ is the 
unique level $n$ FV winner of the election, since $w$ is the only candidate on level 
$n$ with a strict majority of approvals.

From right to left: Suppose $w$ can be made a unique FV winner by adding at most
$k$ candidates denoted by $B'$. By adding candidates from candidate set $B$, only votes in voter
group 1 are changed. Note that candidate $c$ has already $n$ approvals on level
$n-1$ in voter group 2 thus, $c$ can not have any more approvals on level $n$
(else, $\scoresublevel{((C-B)\cup B',V)}{n}{c}\geq n+1$ so, $c$ would tie
or beat $w$ on level $n$). This is possible only if candidate $c$ is pushed 
in all votes in voter group 1 at least one position to the right. This, however,
is possible only if $G$ has a dominating set of size $k$.

For the $\wtwo$-hardness proof in the destructive case, we have to do minor changes 
to the construction, and we will change the roles of candidates $c$ and $w$. Let  
$(G=(B,E),k)$ be a given instance of Dominating Set as described above.
Define the election $(C,V)$, where $C=\{c,w\} \cup B \cup X \cup Y \cup Z$
with $X=\{x_1, x_2, \ldots , x_{n-1}\}$, 
$Y=\{ y_1, y_2, \ldots ,y_{n-2}\}$, $Z=\{z_1, z_2,\ldots ,z_{n-2}\}$ is the 
set of candidates, $c$ is the distinguished candidate, and $V$ is the following
collection of $2n+1$ voters:

\begin{enumerate}
\item For each $i$, $1\leq i \leq n$, there is one voter of the form:
\[
\begin{array}{c@{\ \ }c@{\ \ }c@{\ \ }c@{\ \ }c}
N_c[b_i] & X & c & | & (B-N_c[b_i]) \cup Y \cup Z \cup \{w\}.
\end{array}
\]
\item There are $n$ voters of the form:
\[
\begin{array}{c@{\ \ }c@{\ \ }c@{\ \ }c@{\ \ }c}
Y & c & w & | & B\cup X\cup Z.
\end{array}
\]

\item There is one voter of the form:
\[
\begin{array}{c@{\ \ }c@{\ \ }c@{\ \ }c@{\ \ }c}
Z & w & c & | & B\cup X\cup Y.
\end{array}
\]
\end{enumerate}

Note that again only candidates $c$ and $w$ have strict majority of approvals
in election $(C-B,V)$, both reach the strict majority on level $n$ with   $\scoresublevel{(C-B,V)}{n}{w}=n+1<2n+1=\scoresublevel{(C-B,V)}{n}{c}$ thus,
$c$ is the unique FV winner of the election $(C-B,V)$. Again, let $C-B$ be the 
set of qualified candidates and let $B$ be the set of spoiler candidates.

We claim that $G$ has a dominating set of size $k$ if and only if $c$ can be prevented 
from being a unique FV winner by adding at most $k$ candidates.

From left to right: Suppose $G$ has a dominating set $B'$ of size $k$. Add the corresponding
candidates to the election. Now candidate $c$ gets pushed at least one position to the 
right in each of the $n$ votes in the first voter group. Thus, on level $n-1$ none of 
the candidates has a strict majority of approvals, and 
$\scoresublevel{(C\cup B',V)}{n}{c}=n+1=\scoresublevel{(C\cup B',V)}{n}{w}$, i.e., both
candidates $c$ and $w$ reach a strict majority of approvals on level $n$, and since their 
level $n$ score is equal, $c$ is not the unique FV winner of the election anymore.

From right to left: Suppose $c$ can be prevented of being a unique FV winner by adding at most
$k$ candidates denoted by $B'$. By adding candidates from candidate set $B$, only votes in voter
group 1 are changed. Note that candidate $c$ has already $n+1$ approvals until level
$n$ (including level $n$) in voter groups 2 and 3 thus, $c$ can not have any more approvals on level $n$
(else, $\scoresublevel{((C-B)\cup B',V)}{n}{c}\geq n+2$ so, $c$ would still be the
unique level $n$ FV winner of the election). This is possible only if candidate $c$ is pushed 
in all votes in voter group 1 at least one position to the right. This, however,
is possible only if $G$ has a dominating set of size $k$.~\end{proofs}

\begin{theorem}
 \label{thm:deleting-candidates}
 Both constructive and destructive control by deleting candidates 
in fallback voting are $\wtwo$-hard.
\end{theorem}

\begin{proofs}
We will start with the destructive case. 
Let $(G=(B,E),k)$ be a given instance of Dominating Set.
Define the election $(C,V)$, where $C=\{c,w\} \cup B \cup X \cup Y \cup Z$
with $X=\{x_1, x_2, \ldots , x_{n^2-\sum_{i=1}^{n}||N_c[b_i]||}\}$, 
$Y=\{ y_1, y_2, \ldots ,y_{n-1}\}$, $Z=\{z_1, z_2,\ldots ,z_{n-2}\}$ is the 
set of candidates, $c$ is the distinguished candidate, and $V$ is the following
collection of $2n+1$ voters:

\begin{enumerate}
\item For each $i$, $1\leq i \leq n$, there is one voter of the form:
\[
\begin{array}{c@{\ \ }c@{\ \ }c@{\ \ }c@{\ \ }c}
N_c[b_i] & X_i & w & | & (B-N_c[b_i])\cup (X-X_i) \cup Y \cup Z \cup \{c\},
\end{array}
\]
where $X_i=\{x_{1+(i-1)n-\sum_{j=1}^{i-1}||N_c[b_j]||}, \ldots ,
x_{in-\sum_{j=1}^{i}||N_c[b_j]||} \}$.
\item There are $n$ voters of the form:
\[
\begin{array}{c@{\ \ }c@{\ \ }c@{\ \ }c}
Y &  c & | & B\cup X\cup Z \cup \{ w\}.
\end{array}
\]

\item There is one voter of the form:
\[
\begin{array}{c@{\ \ }c@{\ \ }c@{\ \ }c@{\ \ }c}
Z & w & c & | & B\cup X\cup Y.
\end{array}
\]
\end{enumerate}

Note that candidate $c$ is the unique level $n$ FV winner of the election $(C,V)$, 
since only $c$ has a strict majority of approvals among all candidates on level $n$.

We claim that $G$ has a dominating set of size $k$ if and only if $c$ can be 
prevented of being a unique FV winner by deleting at most $k$ candidates.

From left to right: Suppose $G$ has a dominating set  $B'\subseteq B$ of size $k$. Delete the 
corresponding candidates. Now candidate $w$ gets pushed at least one position to the 
left in each of the $n$ votes in the first voter group. Since candidate $c$ 
gets a strict majority of approvals no earlier than on level $n$ and 
$\scoresublevel{(C-B',V)}{n}{w}=n+1=\scoresublevel{(C-B',V)}{n}{c}$, candidate
$c$ is not the unique FV winner of the resulting election anymore. 

From right to left: Suppose $c$ can be prevented of being a unique FV winner of the 
election by deleting at most $k$ candidates. Observe that only candidate $w$ can 
prevent $c$ from winning the election, since $w$ is the only candidate other than 
$c$ with a strict majority of approvals. In election $(C,V)$, candidate $w$ gets a 
strict majority of approvals no earlier than on level $n+1$, candidate $c$ not before
 level $n$. Candidate
$w$ could only prevent $c$ from winning by getting a strict majority of approvals
no later than on level $n$.  
This is possible only if candidate $w$ is pushed 
in all votes in voter group 1 at least one position to the left. This, however,
is possible only if $G$ has a dominating set of size $k$.

For the $\wtwo$-hardness proof in the constructive case, we have to change one
voter's vote in the above
construction, and we will again change the roles of candidates $c$ and $w$.
Let $(G=(B,E),k)$ be a given instance of Dominating Set.
Define the election $(C,V)$ analogous to the destructive case above with the 
difference that the distinguished candidate is now $w$, and $V$ is the following 
collection of $2n+1$ voters:

\begin{enumerate}
\item For each $i$, $1\leq i \leq n$, there is one voter of the form:
\[
\begin{array}{c@{\ \ }c@{\ \ }c@{\ \ }c@{\ \ }c}
N_c[b_i] & X_i & w & | & (B-N_c[b_i])\cup (X-X_i) \cup Y \cup Z \cup \{c\},
\end{array}
\]
where $X_i=\{x_{1+(i-1)n-\sum_{j=1}^{i-1}||N_c[b_j]||}, \ldots ,
x_{in-\sum_{j=1}^{i}||N_c[b_j]||} \}$.
\item There are $n-1$ voters of the form:
\[
\begin{array}{c@{\ \ }c@{\ \ }c@{\ \ }c}
Y &  c & | & B\cup X\cup Z \cup \{ w\}.
\end{array}
\]

\item There is $1$ voter of the form:
\[
\begin{array}{c@{\ \ }c@{\ \ }c@{\ \ }c@{\ \ }c}
(Y-\{y_1\}) &  c & w & | & B\cup X\cup Z\cup \{y_1\}.
\end{array}
\]

\item There is one voter of the form:
\[
\begin{array}{c@{\ \ }c@{\ \ }c@{\ \ }c@{\ \ }c}
Z & w & c & | & B\cup X\cup Y.
\end{array}
\]
\end{enumerate}

Note that candidate $c$ is the unique level $n$ FV winner of the election $(C,V)$, 
since only $c$ has a strict majority of approvals among all candidates on level $n$.

We claim that $G$ has a dominating set of size $k$ if and only if $w$ can be made 
a unique FV winner by deleting at most $k$ candidates.

From left to right: Suppose $G$ has a dominating set  $B'\subseteq B$ of size $k$. 
Delete the corresponding candidates. Now candidate $w$ gets pushed at least one 
position to the left in each of the $n$ votes in the first voter group. Since 
candidate $c$ gets a strict majority of approvals no earlier than on level $n$ and 
$\scoresublevel{(C-B',V)}{n}{w}=n+2>n+1=\scoresublevel{(C-B',V)}{n}{c}$, candidate
$w$ is the unique FV winner of the resulting election. 

From right to left: Suppose $w$ can be made a unique FV winner of the 
election by deleting at most $k$ candidates. 
Since candidate $c$ already has a strict majority of approvals on level $n$, $w$
has to beat $c$ no later than on level $n$. 
This is possible only if candidate $w$ is pushed 
in all votes in voter group 1 at least one position to the left. This, however,
is possible only if $G$ has a dominating set of size $k$.~\end{proofs}

\subsection{Voter Control}
\label{sec:results:voter-control}

\begin{theorem}
 \label{thm:constructive-adding-voters}
 Constructive control by adding voters in fallback voting 
 is $\wtwo$-hard.
\end{theorem}

\begin{proofs}
Let $(G=(B,E),k)$ be a given instance of Dominating Set.
Define the election $(C,V\cup W)$, where $C=B\cup \{w,x\}$ is the set 
of candidates, $w$ is the distinguished candidate, and $V\cup W$ is the 
following collection of $n+k-1$ voters:

\begin{enumerate}
\item $V$ is the collection of $k-1$ registered voters of the form:
\[
 \begin{array}{c@{\ \ }c@{\ \ }c}
x & | & B \cup \{w\}.
\end{array}
\]
\item  $W$ is the collection of unregistered voters, where for each 
$i$, $1\leq i \leq n$, there is one voter $w_i$ of the form:
\[
\begin{array}{c@{\ \ }c@{\ \ }c@{\ \ }c}
(B-N_c[b_i]) & w & | & N_c[b_i]\cup \{x\}.
\end{array}
\]
\end{enumerate} 

Clearly, $x$ is the level $1$ FV winner of the election $(C,V)$.

We claim that $G$ has a dominating set of size $k$ if and only if $w$
can be made a unique FV winner by adding at most $k$ voters from $W$.

From left to right:
Suppose $G$ has a dominating set $B'$ of size $k$. Add the corresponding 
voters from set $W$ to the election (i.e., each voter $w_i$ if $b_i\in B'$). 
Now there are $2k-1$ registered
voters, thus the strict majority is $k$. Since only candidate $w$
has an overall score of $k$ (all other candidates have an overall 
score of less or equal $k-1$), $w$ is the unique FV winner of the resulting
election.

From right to left:
Suppose $w$ can be made a unique FV winner by adding at most $k$
voters. Denote the set of added voters by $W'$. Note that 
$\scoresublevel{(C,V\cup W')}{1}{x}=k-1$. Since if a candidate has 
a strict majority of approvals on level $1$, he or she is the unique
winner of the election, $k-1$ can not be a strict majority. This is 
only possible, if $||W'||\geq k-1$. If $||W'||= k-1$ then 
$\scoresub{(C,V\cup W')}{w}=k-1$. In this case $w$ has not a strict
majority of approvals, and since  $\scoresub{(C,V\cup W')}{w}=k-1=
\scoresub{(C,V\cup W')}{x}$, candidate $w$ couldn't be made the unique
FV winner of the election. Thus, $||W'||= k$. Note that 
$\scoresub{(C,V\cup W')}{w}=k>k-1 = \scoresub{(C,V\cup W')}{x}$ and $k$
is also a strict majority. Since we could make $w$ the unique FV
winner of the election, none of the candidates in $B$ can be approved
of by each voters in $W'$, otherwise there would exist a candidate $b\in B$
with $\scoresub{(C,V\cup W')}{b}=k$ and $b$ would get the strict majority of
approvals on a higher level than $w$ (since each voter in $W'$ ranks
all the candidates in $N_c[b_i]$ higher than $w$). This is only 
possible if $G$ has a dominating set of size~$k$.~\end{proofs}

\begin{theorem}
 \label{thm:constructive-deleting-voters}
 Constructive control by deleting voters in fallback voting 
 is $\wtwo$-hard.
\end{theorem}

\begin{proofs}
 To prove $\wtwo$-hardness, we provide again a reduction from Dominating Set.
Let $(G=(B,E),k)$ be a given instance of Dominating Set.
Define the election $(C,V)$, where $C=B\cup \{w\}$ is the set of candidates, $w$ is the 
distinguished candidate, and $V$ is the following collection of 
$2n+k$ voters:

\begin{enumerate}
\item For each $i$, $1\leq i \leq n$, there is one voter $v_i$ of the form:
\[
\begin{array}{c@{\ \ }c@{\ \ }c}
N_c[b_i] & | & (B-N_c[b_i])\cup \{w\}.
\end{array}
\]

\item For each $i$, $1\leq i \leq n$, there is one voter of the form:
\[
\begin{array}{c@{\ \ }c@{\ \ }c@{\ \ }c}
(B-N_c[b_i]) & w & | & N_c[b_i].
\end{array}
\]

\item There are $k$ voters of the form:
\[
\begin{array}{c@{\ \ }c}
 | & B\cup \{w\}.
\end{array}
\]
\end{enumerate}

Note that there is no unique FV winner in the above election, each
candidate ties for first place by overall score $n$.

We claim that $G$ has a dominating set of size $k$ if and only if $w$
can be made a unique FV winner by deleting at most $k$ voters.

From left to right:
Suppose $G$ has a dominating set $B'$ of size $k$. Delete the corresponding 
voters from the first voter group (i.e., each voter $v_i$ if $b_i\in B'$). 
Now since there is no candidate
with a strict majority of approvals on any level, and $\scoreof{b_i}\leq n-1$
for each $b_i\in B$ and $\scoreof{w}=n$, $w$ is the unique FV winner
by score of the resulting election.

From right to left:
Suppose $w$ can be made a unique FV winner by deleting at most $k$
voters. Observe that $w$ could only win the election by score, since 
after deleting $k$ voters who disapprove of $w$ we have still $2n$ voters,
and the overall score of $w$ is $n$ thus, not a strict majority.
However, since $w$ was made a unique FV winner by deleting at most 
$k$ voters, $w$ must have a higher overall score than any other
candidate. This is possible only if each of the candidates in $B$ have lost
at least one point. This, however, is possible only if $G$ has a 
dominating set of size $k$.~\end{proofs}

\section{Conclusions and Open Questions}
\label{sec:conclusions}

In this paper we have studied the parameterized complexity of the control problems
for the recently proposed system of {\it fallback voting}, parameterized by the amount of action taken by the chair.  In the case of constructive control, all of the problems are
$\wtwo$-hard.  A natural question to investigate is whether these problems remain 
intractable
when parameterized by both the amount of action and some other measure.
We have shown that all four problems of constructive and destructive control by adding or deleting candidates are hard for $\wtwo$.  What is the complexity when the parameter is both
the amount of action and the number of voters?
We have also shown that both constructive control by adding and deleting voters are hard for $\wtwo$, and that
both destructive control by adding and deleting voters are in FPT.
What is the complexity of constructive control
parameterized by both the amount of action and the number of candidates?

{\small 
\bibliographystyle{alpha}

\bibliography{fv}
}

\end{document}